\input amstex

\baselineskip=24pt\magnification=\magstep1
 \loadmsam \loadmsbm
\input amssym.tex

\centerline{\bf Remarks on an attempted axiomatisation of}
\centerline{\bf Quantum Mechanics, due to Lucien Hardy,}
\centerline{\bf and Ten Theses on Hilbert's Sixth Problem and Quantum Measurement}
\centerline{Joseph F. Johnson}
\centerline{Math Dept., Univ. of Villanova, and CUNY}

This paper has possibly a very different preoccupation than motivated Lucien Hardy in developing his famous paper, `Quantum Theory from Five Reasonable Axioms' (arxiv.org/ quant-ph/0101012).  This paper comes out of a consideration of Hilbert's Sixth Problem (http://euclid.unh.edu/$\sim$ jjohnson/Hilbert.html), the axiomatisation of physics, as focussed by Wigner [8] in a famous discussion of the Problem of Quantum Measurement (arxiv.org/quant-ph/0502124).

Hilbert meant an axiomatisation of the \it physical \rm \ concepts analogously to how he and others axiomatised the mathematical concepts of geometry.  As he and others had eliminated the need for diagrams or intuition from plane geometry, so he hoped mathematicians could eliminate the need for physical intuition from physics.  From this standpoint, which is one close to Wigner's (or at least he shows a sensitivity and awareness of it), Hardy's so-called axiomatisation is quite alien.  We will focus on what Hilbert himself singled out as fundamental, the concept of physical probability.

This concept was salient in Hilbert's time, as the controversy between Boltzmann and Zermelo showed, but has become even more foundational in physics since 1925.

ON page 10, Hardy re-states his first axiom, which we quote:

{\baselineskip=12pt\narrower\smallskip\noindent `Axiom 1 Probabilities.  Relative frequencies (measured by taking the proportion of times a particular outcome is observed) tend to the same value (which we call the probability) for any case where a given measurement is performed on an ensemble of $n$ systems prepared by some given preparation in the limit as $n$ becomes infinite.
\smallskip}

What I want to claim here is more radical than any comment that he has chosen the `wrong' interpretation of probability, or the `wrong' axiomatisastion of Quantum Mechanics.  This is simply not an axiom at all.  (Dirac and von Neumann are much closer to Hilbert's spirit in this regard: at least each axiom of theirs, taken individually, is an axiom, even though when taken collectively, there are subtle problems which Wigner was aware of.)
He has not specified which are the undefined, primitive terms, and which are the constraints being imposed on the usage of those terms by the axioms.

Is this a definition of probability?  The term ``limit'' is undefined so it is not completely spelled out.  In neither mathematics, nor physics, does the word `limit', by itself, have a definite meaning.

Is this a physical axiom?  It could, at first sight, be a physical axiom that $$ \lim _ {T\rightarrow\infty}\frac1T\int_ 0^Tf_t(v_o)dt$$ $$ {\text {always exists---or}} \lim _ {T\rightarrow\infty}\frac1T\sum_{t=0}^{T-1} f_{v_0} (t),$$ as the case may be---and is, by definition, then set equal to the  notion ``probability.''  It has not, up to now, been thought that in the real world this limit did always exist.  But let us adopt this as an axiom anyway.

Here, $f_{v_0}$ is a sequence (or in general a continuous function).  Its definition is $f_{v_0}=1$ if the  $n^{th}$ trial  is success, 0 if failure.  (There is an issue of constructionism here which is irrelevant).  If $f_{v_0}$ is not a deterministic function of $t$ (and ${v_0}$ the initial condition) then Hardy has not defined ``limit.''  If it is one of the usual definitions of `limit', $f_{v_0}$  must be a function in the usual sense of pure mathematics, a subset of $V\times \Bbb N\times \{0,1\}$ such that $\pi_1(x)= \pi_1(y)$ and $\pi_2(x)=  \pi_2(y)$ implies $\pi_3(x)=  \pi_3(y)$ where $\pi_i$  means projection onto the $i^{th}$ factor.  If it isn't, Hardy has totally missed the point of axiomatics.  

So let us assume that the axiom states that such a function $f$ exists.  That is, whenever the physical set up he mentions exists, then that physical set up yields a function $f_{v_0}:\Bbb N\rightarrow\{0,1\}$ as above, and then the limit exists mathematically and agrees with the mathematical notion of probability.  

There are several major difficulties with this.  Firstly, logic gates.  It must be possible to combine the results of two independent sequences of measurements $f_{v_0}^1,f_{w_0}^2$ into one sequence $g_{v_0,w_0}$ by means of logic gates, for example $g_{v_0,w_0}(n)=f_{v_0}^1(n)\and f_{w_0}^2(n)$. This is a physically realisable measurement.  But even if $f_{v_0}^1$ and $f_{w_0}^2$ satisfy the axiom, $g_{v_0,w_0}$ need not.  (No amount of noise will mask this effect if the noise also obeys the axiom.)   Secondly, his other axioms do not allow us to say that $f_{v_0}$ exists.  (Most interpretations rule out the existence of $f_{v_0}$.)  %So his formal system does not give us any properties of $f$ except the existence of this limit.

Hence $f_{v_0}$ cannot be a function even for given ${v_0}$ (except in special cases).  But then the notion of ``limit'' is undefined.  Thirdly, in mathematics probability is modeled by a measure, or else he should propose an alternative.  He can't say what space his physical probability is a measure on, even in the classical physics case!  %Because of the problem of logic gates,
the space cannot be the space of all mathematical sequences, and he has not defined any other space.

Hardy exhibits a deep (although typical) conceptual confusion over whether he is axiomatising physical probability or mathematical probability.  These are two separate problems even though once one is finished solving each one separately there remains the last task of showing that the one is a model for the other.  

Mathematical statements are only a model for physical statements.  For example, `3' is not a primitive concept in mathematics, it is the set of all sets equipotent to \dots etc.   But the physical concept of `3' would, if axiomatised, not have the same logical structure at all.  Then one checks that the physical axioms imply that then mathematical axioms are true when given the physical interpretation of the model. 

The problem even for the limited context of classical mechanics and Kolmogoroffian probability was not really touched until von Plato's [7] important \it Ergodic Theory of Probability\rm, of 1988, and finished by me in 1990 with \it The Logical Structure of Probability Assertions\rm\ (arxiv.org/quant-ph/0508059).  This last extends to non-Relativistic Quantum Mechanics and any finite or renormalisable Quantum Field Theory.  

Kolmogoroff, indeed, realised and admitted[4] it had not been touched.  Hence his work of the  `60's on algorithmic complexity as a definition of randomness.  But neither he nor his followers made a logically specific connection with physics so it remains a project rather than a theory.  There is no rival, yet, to von Plato's (and my) theory.  

Assuming that the mathematical axiomatisation of classical probability theory is the measure-theoretic one developed by Kolmogoroff, one would have to ask of the physical conception of probability of Hardy's, 
how can the space (where the measure lives on) be defined? Is it the space of all (mathematical) sequences of results of measurements?  Or only those which are `physical' in some vague ill-defined sense?  The \it whole \ \rm task is to make that sense precise.  (The thermodynamic limit does this.)
It can't be the space of all sequences of observations, since that space, although 
a measure space, possesses an element, i.e., a sequence, not having a frequency.
It can't be the subspace of that space consisting of all those sequences possessing 
frequencies, since that space is, as we remarked, obviously not closed under simple 
physically implementable combinations of observations.  %(All of this is well-known.)

\vskip -8pt
Could he try to rule out the counter-examples under discussion by imposing a physical
condition?  (This would be analogous to Kolmogoroff's attempt to impose a condition on 
the sequences by imposing a logical constructibility condition on the algorithm which 
would construct the sequence.)
This is no easy task, as the vicissitudes of Kolmogoroff's project showed.
It would necessitate making the sequence of observations a determined function of a 
physical specification of the experimental set-up, instead of `random.'  (Or else specifying a new definition of limit instead of using one of the normal ones: this in fact is what is done in the ergodic theory of probability and in its generalisation.)
It would necessitate subsuming both measurement and transformation under one over-arching
concept of `physicality', i.e., dealing with Wigner's problem and Bell's objection [1].
(http://euclid.unh.edu/$\sim$ jjohnson/Bellagainst.html).
The whole idea of random variable shows this: its logical structure is, unfortunately, that  the event is a deterministic function 
of something\dots an element of a measure space, a hidden variable.  
If he insists on using the usual notion of limit of a function (or sequence), then
the physics must specify a mathematical object which is a function (or sequence), and 
he can't do this.  IMHO, it is preferable to abandon that road and instead
 define by physical procedures a 
measure of total mass one in some other specified way.  It does not really seem practical to model the idea of a sequence of measurements by a mathematical sequence.  It is necessary to have the physical specification of that measure agree with experiment in the sense that the calculations, using the mathematical theory of probability, performed on that measure, agree to within experimental error with the observed probabilities which form the backbone of the experimental support for Quantum Theory.  But we would not need to have a naive structure of `events' and `outcomes.'

TEN THESES ON THE AXIOMATISATION OF PHYSICS

1.The experimental evidence, even today, still points to Dirac's axioms, except 
possibly the reduction of the wave packet.

%(The only new evidence about Quantum Mechanics not available to him in 
%1930 is that a) parity is not conserved\dots now his book carefully refrains 
%from saying that it is, and b) macroscopic entanglement is possible, \dots 
%now his book carefully allows Schroedinger's equation, and hence 
%entanglement, to have universal validity, including the macroscopic 
%region.  The reduction of the wave packet is not directly supported by evidence,
%but every statistical correlation which can be deduced from it, is so supported.)

2.The salient defect of these axioms is merely what Wigner pointed to, 
their dual overlap: a measurement can be analysed in two seemingly incompatible ways.

3.The foundations of physical probability are in a logically unsatisfactory 
state, so it is at least conceivable that fixing this could be part of the solution.

4.The frequency theory of probability is logically circular, as is well 
known (Kolmogoroff[4], Burnside[2], Littlewood[5]).  But it is so well-supported 
experimentally that what is needed is a logically unexceptionable
theory of the physical meaning of a probability assertion that remains
as close as possible to the frequency theory.  Von Plato has provided
one, and his suggestion can be improved.

5.Statistical Mechanics, in its logically careful formulation, does not 
use open systems, coarse-graining, `information', infinite systems, or entropy.

6.Dirac's axioms do not allow one to talk about open systems except in the 
same way one was accustomed to do in classical physics (as Hamiltonian heat baths).

7.A concept, such as mixed state or density matrix, can be rigourously 
defined in terms of the primitives of Dirac's axiom system, but it does not thereby 
acquire any independent physical meaning.  Since logical definitions are 
merely abbreviative, saying `mixed state' is merely an abbreviation for 
a complicated logical assemblage of statements about pure states, 
probabilities, coefficients, linear combinations, etc.  %As in Russell's 
%Principles of Mathematics.
(Russell [6],  p.\ 4.)

8.It behooves us to investigate whether or not the puzzle can possibly be 
solved without otherwise unmotivated alterations in the terms of the 
problem first, before falling back on changing the terms of the problem.
(Non-linearity, open systems, hidden variables, information, coarse-graining, 
etc.)

9.An analysis of the physics of actual measurement apparatuses, e.g., 
amplification, would be likely to point the way to an answer.  It would 
seem natural to suppose that the phenomena of measurement depend on 
the physics of measurement apparatuses!  (Either that, or on the 
physics of interaction with the environment, but that should be a fall 
back in case the analysis of the physics of amplification is unsuccessful.)

10.A priori philosophical assumptions should not be allowed much weight 
against a theory that agrees with experiment.  These include interpretations 
of physical probability, psycho-physical parallelism, monism, etc.
On the other hand, J.S. Bell's requirements of logical coherence, stating 
clearly what is assumed and distinguishing it from what is proved, are not 
a priori philosophical assumptions, they are criteria needed before one 
even *had* a theory to compare with experiment at all in the first place.

Experimental result means *replicable*.  Something experienced does not count 
as an experimental result unless it is replicable.  Nature and Heisenberg 
have taught us that only the probabilities of our subjectively experienced 
conscious meter-readings are replicable, therefore only the probabilities 
are experimental results, and a theory that disagrees with our conscious 
perceptions does not thereby disagree with experiment.

%\vfill \eject
\centerline{\bf Bibliography}
\noindent [1] J.\ Bell, Phys. World 3 (1990), 33.

\noindent [2] W.\ Burnside,  ``On the Idea of Frequency,'' Proc.\ Camb.\ Phil.\ Soc.\rm, \bf 22 \rm (1925), 726.

\noindent [3] J.\ Johnson, in \it Quantum Theory and Symmetries III\rm, Cincinnati, 2003, ed.\ by Argyres \it et al\rm., Singapore, 2005, 133.

\noindent [4] A.\ Kolmogoroff, in \it Mathematics its Content, Methods, and Meaning\rm, ed.\ by Aleksandroff, Kolmogoroff, and Lavrentieff, Moscow, 1956 p.\ 239. 

\noindent [5] J.\ Littlewood, \it A Mathematician's Miscellany\rm, London, 1953, p.\  55f.

\noindent [6] B.\ Russell, \it Introduction to Mathematical Philosophy\rm, London, 1919.

\noindent [7] J.\ von Plato, Ergodic Theory and the Foundations of Probability, in Brian Skyrms and William Harper (eds.), Causation, Chance, and Credence, Proceedings of the Irvine Conference on Probability and Causation, vol.\ 1, pp.\ 257-277, Kluwer, 1988.

\noindent [8] E.\ Wigner, Am.\ Jour.\ Phys.\ 31 (1963), 6.

Several important papers have been written which accept Thesis 9.
However, they are far from the spirit of Hilbert's problem.  In 
fact, for the reason J.\ S.\ Bell pointed to, they are anti-axiomatic.
It is anti-axiomatic to pass from a density matrix with vanishing
off-diagonal terms to a probability interpretation, or, as Bell 
puts it, `to pass from an *and* to an *or*.'
`The idea that elimination of coherence, in one way or another, 
implies the replacement of ``and'' by ``or'',
is a very common one among solvers of the ``measurement 
problem''.  It has always puzzled me.' (\it loc.\ cit\rm.)
By elimination of coherence, for example, he refers to the 
disappearance of off-diagonal terms in the density matrix.
Under Schroedinger's influence, H.\ S.\ Green wrote Nuovo Cimento {\bf9} (1958), 880,
which treats a negative temperature heat bath as making a transition
under a microscopic stimulus.  Daneri--Loinger--Prosperi, Nucl.\ Phys.\ {\bf33} (1962), 297, analysed 
in general a similar situation. 
K. Hannabuss, Helv.\ Phys.\ Acta {\bf57} (1984), 610; Ann.\ Phys.\ (NY) {\bf239} (1995), 296 is more explicit still.
Allahverdyan--Balian--Nieuwenhuizen (cond-mat/0102428)
have explicitly analysed a driven phase transition as a model of the 
measurement process.  These treatments all transgress axiomatic
procedures in general and theses 3, 5, and 7 in particular, which is alright 
since they are physics papers, not purporting to be axiomatic or solve a Hilbert 
problem.
Johnson (2003) quant-ph/0502044 is similar, in underlying physical mechanism, to 
the above, but is axiomatic in spirit as Wigner was in his posing of the problem 
of Quantum Duality (and has a different explicit definition of macroscopic pointer variable).
\end